# Computational impact of hydrophobicity in protein stability

Geetika S. Pandey[1]
Research Scholar,
CSE dept., RGPV,
Bhopal (M.P), India
geetika.silakari@gmail.com

Dr. R.C Jain[2]
Director, SATI(D),
Vidisha(M.P), India
dr.jain.rc@gmail.com

*Abstract-* **Among the various features of amino acids, the hydrophobic property has most visible impact on stability of a sequence folding. This is mentioned in many protein folding related work, in this paper we more elaborately discuss the computational impact of the well defined 'hydrophobic aspect in determining stability', approach with the help of a developed 'free energy computing algorithm' covering various aspects - preprocessing of an amino acid sequence, generating the folding and calculating free energy. Later discussing its use in protein structure related research work.**

## I. INTRODUCTION

Since the earliest of proteomics researches, it has been clear that the positioning and properties of amino acids are key to structural analysis [1]. According to Betts et.al. in the protein environment a feature of key importance is cellular location. Different parts of cells have very different chemical environments with the consequence that many amino acids behave differently. The biggest difference as mentioned by Betts et.al. is between soluble proteins and membrane proteins. The soluble proteins tend to be surrounded by water molecules i.e have polar or hydrophilic residues on their surface whereas membrane proteins are surrounded by lipids i.e they tend to have hydrophobic residues on the surface that interact with the membrane. Further the soluble proteins are categorized as extracellular and intracellular. So basically through the various studies [2] could conclude that the core of protein contains hydrophobic amino acids forming certain bonds and thus structures. the stability of the structures is determined by the free energy change, as mentioned by Zhang et. al [3] i.e.

$\Delta G(folding) = G(folded) - G(unfolded)$ [3]

Later in this paper various aspects of folding and stability are discussed in detail.

## II. BACKGROUND

### A. Features

Shaolei Teng et.al.[4] mentioned twenty amino acid features which they used to code each amino acid residue in a data instance. They obtained these features from Protscale (http://expasy.org/tools/protscale.html) [5] and AAindex (http://www.genome.jp/aaindex/) [6]. They further mentioned these features into four categories -

Biochemical features – includes M, molecular weight, this is related to volume of space that a residue occupies in protein structure. K, side chain pka value, which is related to the ionization state of a residue and thus plays a key role in pH dependent protein stability. H, hydrophobicity index, which is important for amino acid side chain packing and protein folding. The hydrophobic interactions make non-polar side chains to pack together inside proteins and disruption of these interactions may cause protein destabilization. P, polarity, which is the dipole-dipole intermolecular interactions between the positively and negatively charged residues. Co, overall amino acid composition, which is related to the evolution and stability of small proteins.

<u>Structural features</u>- this includes A, alpha-helix. B, beta-sheet. C, coil. Aa, average area buried on transfer from standard state to folded protein. Bu, bulkiness, the ratio of the side chain volume to the length of the amino acid.

<u>Empirical Features</u>- this includes, S1, protein stability scale based on atom atom potential of mean force based on Distance Scaled Finite Ideal-gas Reference (DFIRE). S2, relative protein stability scale derived from mutation experiments. S3, side-chain contribution to protein stability based on data from protein denaturation experiments.

<u>Other biological features</u>- F, average flexibility index. Mc, mobility of an amino acid on chromatography paper. No, number of codons for an amino acid. R, refractivity, protein density and folding characteristics. Rf, recognition factor, average of stabilization energy for an amino acid. Rm, relative mutability of an amino acid. Relative mutability indicates the probability that a given amino acid can be changed to others during evolution. Tt, transmembrane tendency scale. F, average flexibility index of an amino acid derived from structures of globular proteins.

### B. Protein folding

Protein folding has been considered as one of the most important process in biology. under the various physical and chemical conditions the protein sequences fold forming bonds , when these conditions are favourable the folding leads to proper biological functionality. But some conditions could lead to denaturation of the structures thus giving unfolded structures. protein denaturants could be [7] –

- High temperatures, can cause protein unfolding, aggregation.
- Low temperatures, some proteins are sensitive to cold denaturation.
- Heavy metals(e.g. lead, cadmium etc), highly toxic, efficiently induce the 'stress response'.
- Proteotoxic agents(e.g. alcoholc, cross-linking agents etc.)
- Oxygen radicals, ionizing radiation- can cause permanent protein damage.
- Chaotropes (urea, guandine hydrochloride etc.), highly potent at denaturing proteins, often used in protein folding studies.

Protein folding considers the question of how the process of protein folding occurs, i.e how the unfolded protein adopts the native state. Very often this problem has been described as the second half of the genetic code. Studies till date conclude the following steps as the solution for this problem [8] –

- 3D structure prediction from primary sequence.
- Avoiding misfolding related to human diseases.
- Designing proteins with novel functions.

### C. Factors affecting protein stability

Protein stability is the net balance of forces which determine whether a protein will be in its native folded conformation or a denatured state. Negative enthalpy change and positive entropy change give negative i.e. stabilizing, contributions to the free energy of protein folding, i.e. the lower the $\Delta G$, the more stable the protein structure is [7]. Any situation that minimizes the area of contact between $H_2O$ and non-polar, i.e hydrocarbon, regions of the protein results in an increase in entropy [9].

$$\Delta G = \Delta H - T\Delta S$$

Following are the factors affecting protein stability [8]:
- pH : proteins are most stable in the vicinity of their isoelectric point, pI. In general, with some exceptions, electrostatic interactions are believed to contribute to a small amount of the stability of the native state.

- Ligand binding: binding ligands like inhibitors to enzymes, increases the stability of the protein.
- Disulphide bonds: it has been observed that many extracellular proteins contained disulphide bonds, whereas intracellular proteins usually did not exhibit disulphide bonds. Disulphide bonds are believed to increase the stability of the native state by decreasing the conformational entropy of the unfolded state due to the conformational constraints imposed by cross linking (i.e decreasing the entropy of the unfolded state).
- Dissimilar properties of residues: not all residues make equal contributions to protein stability. Infact, studies say that the interior ones, inaccessible to the solvent in the native state make a much greater contribution than those on the surface.

## III. EXPERIMENTAL PROCEDURE

### A. Approach

As per the amino acid features mentioned previously, the hydrophobic property is most responsible for the folding, as well as stability related issues. Hence in the algorithm mentioned later this property is taken as the key in preprocessing of the input sequence, i.e. the binary representation where '1' denotes the hydrophobic amino acids and others as '0', as per the hydrophobicity scales proposed by Kyle et. al [9]. Then using the complex plane the folding configurations are formed and their combinations denote various turns [10]. The cumulative sum of the configuration is calculated which gives the direction of each fold. Later the free energy of each folding is calculated using Euclidean distance between the hydrophobic amino acids i.e. all 1s and as per the study the folding having lower free energy value would be stable hence the stable structures could be obtained.

### B. Data

The data in this case is a protein sequence loaded from protein data bank with pdb id 5CYT, heme protein, using Matlab 7.

Pro= 'XGDVAKGKKTFVQKCAQCHTVENGGKHKVGPNLWGLFGRKTGQAEGYSYTDANKSKGIVWNNDTLMEYLENPKKYIPGTKMIFAGIKKKGERQDLVAYLKSATS'

### C. Methods

In brief the steps are as follows:

1) Preprocessing of the input primary protein sequence using the hydrophobicity scale developed by Kyte & Doolittle [9], i.e. developing a vector with hydrophobic amino acids represented by 1 and hydrophilic by 0.
2) Calculating the free energy of this initial sequence
3) Now generating various foldings through iteration, using complex number 'i'.
4) Calculating the free energy for all these foldings.
5) Now further these free energy values could be used to check the stable structures.

### D. Algorithm

Input – an amino acid sequence, Pro.

Output- an array of free energy of each structure predicted, E.

1) Preprocessing of the input protein sequence
   a) N ← length(Pro)
   b) bin ← Pro
   c) for idx ← 1:N
   d) if Pro(idx)= hydrophobic
   e) then bin(idx)← 1
   f) else bin(idx) ← 0
   g) end
   h) end
2) folding formation
   a) conf ← ones(length(bin)-1,1)
   b) e ← Free_energy(conf)
   c) for k ← 2:length(conf)
   d) f(1:k) ← i
   e) f(k+1:end)←−1

  f) conf ← conf*f
  g) F(:,count) ← conf
  h) count = count+1
  i) end
3) free energy of all the structures in F(m,n)
  a) for j ← 1:n
  b) q ← F(:,j)
  c) p ← Cumulative_sum(q)
  d) E(j) ← Free_Energy(p)
  e) End
4) Algorithm for Cumulative Sum
  Cumulative_sum (a)
   a) for x ← 1 : length(a)
   b) sum ← sum + a(x)
   c) end
5) Algorithm for Free energy
  Free_Energy(a)
   a) a ← a * (bin with only hydrophobic elements)
   b) for x ← 1 : length(a)
   c) d ← abs( a(x) –a(x+1))
   d) sum ← sum + d
   e) end
   f) energy ← sum

## IV. Results

The length of the sequence in this case was 104, hence as the algorithm total number of folding created is 103, each column of matrix F (fig. 1) shows a folding. And each row of array E (fig. 2) shows the free energy for each folding. Here the free energy of the unfolded structure is 'e= 45194'.

## V. Discussion and futurework

The result from this approach provides the practical aspect of the impact of hydrophobicity on stability, the various outcomes could be used for further research or with some modifications could lead the ultimate solution. With the help of this method the folding could be generated at any structure level, these folding could be used for further research work like in machine learning or neural networks. The free energy calculated could be further used for clustering or classification purposes, thus could enhance the study of the stability factors. In the future work hydrophobicity could be coupled with any other amino acid feature.

AUTHORS PROFILE

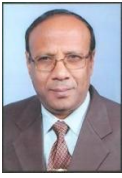 R. C. Jain, M.Sc., M. Tech., Ph. D., is a Director of S.A.T.I. (Engg. College) Vidisha (M. P.) India. He has 37 years of teaching experience. He is actively involved in Research with area of interest as Soft Computing, Fuzzy Systems, DIP, Mobile Computing, Data Mining and Adhoc Networks. He has published more than 125 research papers, produced 7 Ph. Ds. and 10 Ph. Ds are under progress.

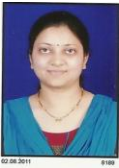 Geetika S. Pandey obtained her B.E degree in Computer Science and Engineering from University Institute of Technology, B.U, Bhopal in 2006. She obtained Mtech degree in Computer Science from Banasthali Vidyapith, Rajasthan in 2008. She worked as Assistant Professor in Computer Science and Engineering Department in Samrat Ashok Technological Institute, Vidisha (M.P). She is currently pursuing Ph.D. under the supervision of Dr. R.C Jain, Director, SATI, Vidisha. Her research is centered on efficient prognostication and augmentation of protein structure using soft computing techniques.

# Appendix

Fig. 1, F(103x103), various folding of sequence pro.

Fig. 2, E(103x1), free energy of each folding.